\documentclass[11pt,a4paper,twoside]{article}
\usepackage{changepage} 
\usepackage{graphicx}
\usepackage{verbatim}
\usepackage{amsmath,amssymb,amsfonts} 
\usepackage{bm}

\usepackage[T1]{fontenc}
\usepackage{lmodern}

\usepackage{titlesec}

\titleformat{\section}{\normalfont\Large\color{black}}{\bf\thesection}{1em}{}

\titleformat{\subsection}{\normalfont\large\color{black}}{\bf\thesubsection}{1em}{}

\usepackage[a4paper,inner = 2.5cm, outer = 3.0cm, top = 2.5cm, bottom = 2.5cm]{geometry} 
\usepackage{fancyhdr}
\fancypagestyle{plain}{ %
  \fancyhf{} 
  
}

\usepackage{epstopdf}
\DeclareGraphicsExtensions{.png,.jpg,.jpeg,.eps}
\usepackage{float}
\usepackage{caption}
\usepackage{subcaption}

\usepackage{cite}

\usepackage{siunitx}
\usepackage{authblk}

\title{High spectro-temporal purity single-photons from silicon micro-racetrack resonators using a dual-pulse configuration}

\author[1,2,$\dagger$]{Ben Burridge}
\author[1,$\dagger$]{Imad I. Faruque}
\author[1]{John Rarity}
\author[1,2]{Jorge Barreto\thanks{G.Barreto@bristol.ac.uk}}

\affil[1]{Quantum Engineering Technology Laboratories, University of Bristol, Bristol, United Kingdom.}
\affil[2]{Quantum Engineering Centre for Doctoral Training, Centre for Nanoscience \& Quantum Information, University of Bristol, Bristol, United Kingdom.}
\affil[$\dagger$]{These authors contributed equally}

\date{}                     

\begin{document}
\maketitle
\begin{abstract}
Single-photons with high spectro-temporal purity are an essential resource for quantum photonic technologies. The highest reported purity up until now from a conventional silicon photonic device is 92\% without any spectral filtering. We have experimentally generated and observed single-photons with 98.0 $\pm$ 0.3 \% spectro-tempral purity using a conventional micro racetrack resonator and an engineered dual pump pulse.
\end{abstract}

Quantum information technologies promise unbreakable secure communications, novel sensing, detection free imaging, and fast computations beyond what is achievable using conventional technologies \cite{Shor1999, Knill2001, Raussendorf2001, Gottesman1999, Ekert1992, Brida2010, Lemos2014, Wang2018}. In the last few years heroic proof-of-principle experiments have demonstrated the foundations of quantum technologies, primarily using bulk optical setups \cite{Ekert1992, Brida2010, Lemos2014, Wang2018}. Real-world applications require scaling up these demonstrations to 10 - 30 photon experiments in the near term, which will be unreasonably challenging due to the space and stability constraints of bulk optics. Integrated optics have been used to demonstrate larger and more sophisticated quantum photonic experiments \cite{WangJ2019}, many of them specifically on a silicon photonics platform: arbitrary two quantum bit (qubit) processing \cite{Qiang2018}; the first chip-to-chip quantum teleportation \cite{LLewellyn2020} and state-of-the-art multi-dimensional Bell inequalities \cite{WangJ2018}. Utilising the fabrication technologies developed initially for micro-electronics applications, silicon photonics has demonstrated the potential to realise these quantum technologies at a commercial scale using low-cost solutions, and with unparalleled fabrication precision and electronic integration \cite{Abel2018, Sun2015}. The primary promise of the silicon photonics platform is its scalability: to deliver large and complex architecture (i.e. small device footprint) achieving the desired quantum advantage \cite{Rudolph2017, Silverstone2016}. 

One of the central challenges for realizing scalable silicon quantum photonic devices is to develop high-performance sources \cite{Rudolph2017}. Single-photon sources need to produce simultaneously photons with near unity spectro-temporal purity and near unity photon-number purity while also being bright, near deterministic, and single-moded in all aspects. For scalability proposes, different sources have to be highly (>99\%) indistinguishable with respect to each other \cite{Rudolph2017}. The most common structures used in silicon photonics as sources are long waveguides and micro-resonators \cite{Qiang2018, LLewellyn2020, WangJ2018}; these structures are used to exploit a nonlinear optical process, spontaneous four-wave mixing (SFWM). In SFWM, pump pulses generate pairs-of photons; signals and idlers, as they propagate through waveguides or micro-resonators conserving energy and momentum. We can herald the presence of single signal photons upon detecting the presence of idler photons, thus effectively acting as a single-photon source. Spectral filtering has been used to increase the spectro-temporal purity and the indistinguishability of the waveguide sources at the cost of reduced heralding efficiency, or brightness of the sources \cite{Blay2017, Faruque2019, Zhang2016a}. In contrast, without any spectral filtering, conventional micro- ring or micro-racetrack resonators (MRRs) are fundamentally bound to a high but maximum purity of 92\% \cite{Helt2010, Grassani2016, Faruque2018}, and likewise similar values of indistinguishability \cite{LLewellyn2020}.

\begin{figure*}[ht]
\centering
\includegraphics[width=\linewidth]{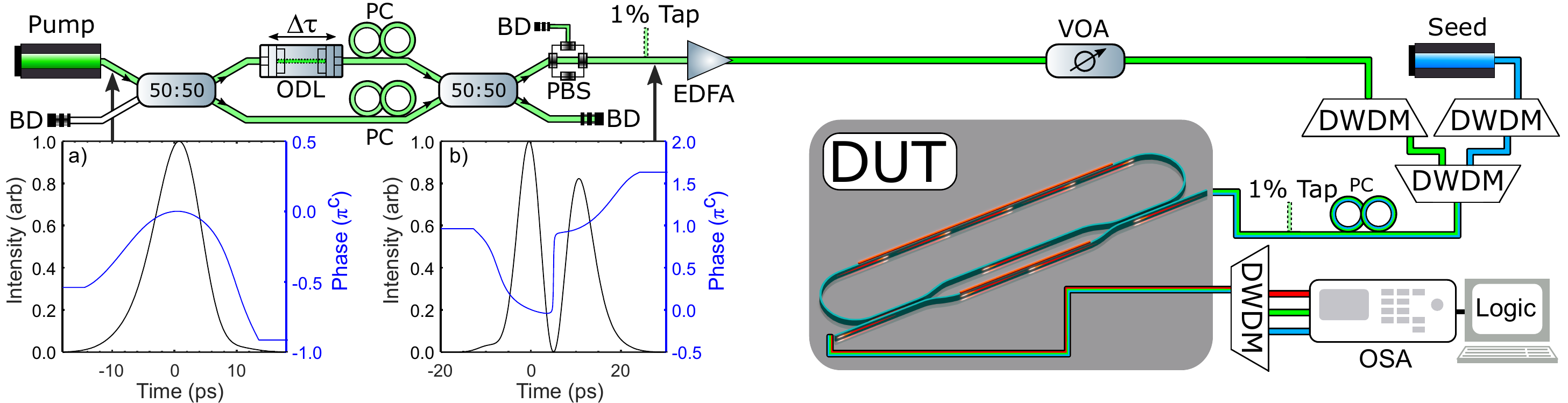}
\caption{A schematic of the dual-pulse construction and measurement of spectro-temporal purity. Top left shows the interferometric optical setup for the dual pump pulse construction. Bottom left contains the single (a) and dual pump (b) pulse with required $\pi$ phase shift, as measured using a FROG. The right side of the figure shows the tomography on our MZI-coupled MRR to measure the JSI and estimate the purity. Here, BD: Beam Dump; PC: Polarisation Controller; ODL: Optical Delay Line; PBS: Polarisation Beam Splitter; EDFA: Erbium doped fibre amplifier; VOA: Variable Optical Attenuator; DWDM: Dense Wavelength Division Multiplexer; OSA: Optical Spectrum Analyser.}
\label{fig:DUT}
\end{figure*}

Two different theoretical proposals suggest that it is possible to achieve arbitrarily high (>99\%) spectro-temporal purity by either engineering the MRR couplings (e.g. interferometrically) \cite{Vernon2017b, Gentry2016} or by manipulating the pump pulse \cite{Christensen2018}. Both methods can be interpreted as an effective broadening of the pump resonance compared to the signal-idler resonances, resulting in an increase in purity.
Recent experiments \cite{Liu2020, Faruque2020} have verified an increased spectro-temporal purity of the single-photons generated from the interferometrically coupled and complex resonant structures proposed in \cite{Vernon2017b}. These structures have footprints larger than conventional MRRs, the added complication of multiple phase shifters, and their use applicable mainly to photon-pair generation.
The second proposal \cite{Christensen2018} relies on using conventional MRRs while instead manipulating the pump pulse in a time-delayed dual-pulse configuration. A key benefit of using this dual-pulse method is that the purity of existing resonator structures can be easily increased using existing equipment that is readily available. Predicted purities are expected to be well in excess of the limited 92\% purity achievable in the single pulse regime. This limitation comes from the identical resonances of conventional ring resonators, restricting the linewidth of the in-resonator pump spectrum to that of the pump resonance. Using two pulses allows us to tune the temporal response of the resonator, effectively broadening the in-resonator pump spectrum compared to that of the signal / idler fields. Here we have constructed a dual pulse from the original pump pulse and estimated the spectro-temporal purity of the single-photons using joint spectral intensity (JSI) measurements.


A PriTel FFL pulsed laser with 9 ps (420 pm) or 16.7 ps (280 pm) pulse widths, and a 50 MHz repetition rate is used as a pump source in our experiment. The experimental setup is schematically shown in Fig.~\ref{fig:DUT}.
We have constructed the dual-pulse using an off-chip asymmetric Mach-Zehnder interferometer (AMZI). Starting from the top-left part of the figure, the pulses from the pump laser are incident on a 50:50 fiber-optic beam splitter and divided in two paths. One of the pulses propagates through one of the output arms of the splitter which is connected to an optical delay line (ODL) and then to a polarisation controller (PC), while the other pulse propagates through the other arm and then to another PC. The outputs of the two PCs are combined in another 50:50 fibre-optic beam splitter and then to a fibre-optic polarisation beam splitter (PBS), therefore, allowing us to dynamically control the splitting ratio ($\eta$) between the two pulses. The PBS also serves to restrict the output of the AMZI to a single polarisation mode, maximising interference visibility. The ODL temporally delays the propagation in one arm of the interferometer with respect to the other, effectively resulting in two pulses. The exact amount of delay ($\Delta\tau$ - relative to the pulse width $\tau_{p}$) is dependent on the central frequency ($\nu_p$) of the pump resonance of the MRR, such that the delay results in a $\pi$ phase shift centered at this frequency. The phase and time delay of the constructed dual-pulse compared to the original single pulse are estimated using Frequency-Resolved Optical Gating (FROG), as shown in Fig.~\ref{fig:DUT} a) \& b). 
By expressing the optical frequency as $\nu$ and the pump pulse using sech, the resultant pulse can be described as: 
\begin{align}
\alpha_p(\nu) = \left[\sqrt{\eta} - \sqrt{1-\eta}\exp{\big(-i\Delta\tau\left(\nu-\nu_0\right)\big)}\right]  {\mathrm{sech}{\big(\left(\nu-\nu_p\right)\tau_p\big)}}
\end{align}
\label{eq:dual_pump}
A PriTel FA-20 Erbium-doped fibre amplifier (EDFA) is positioned after the AMZI to strengthen the resultant pump pulse, followed by a Variable Optical attenuator (VOA) to finely tune the power into the device. It has to be said that the AMZI described here is entirely implemented using off-the-shelf fibre-optics, and both passive thermal and vibration isolation strategies are required to minimise interferometer drift with time. 

We have implemented a stimulated emission tomography (SET) \cite{Liscidini2013} method to measure the JSI to estimate the Joint Spectral Amplitude (JSA). The JSA captures the energy and momentum conservation of the SFWM photon-pairs generated from our MRR, and is usually expressed by the following equation \cite{Helt2010, Vernon2017b, Christensen2018},
\begin{align}
f(\nu_s, \nu_i) = \int d\nu_p~& \alpha_p(\nu_p)\alpha_P(\nu_s + \nu_i - \nu_p) \phi(\nu_s, \nu_i) L(\nu_p)  L(\nu_s + \nu_i - \nu_p) L(\nu_s)^*L(\nu_i)^*  
\label{eq:JSA}
\end{align}
where, $\nu_s$ \& $\nu_i$ are the frequencies of generated signal \& idler photons, $\phi(\nu_s, \nu_i)$ is the momentum conservation or phase matching, and $L(\nu)$ is the Lorentzian linewidth field-enhancement of the MRR. The phase-matching function is conserved and assumed to be unity in the telecom wavelength for silicon photonics. The Schmidt decomposition of the $f(\nu_s, \nu_i)$, that is singular value decomposition of the estimated JSA from the JSI, gives us the spectro-temporal purity value \cite{Grassani2016}.

 In the SET of SFWM, a seed laser is used to represent the signal photons ($\nu_s$), and stimulated idler photons are generated at $\nu_i$ frequencies. The seed laser used here is a Yenista Tunics HPS with a resolution between 2-4 pm. We have used Dense Wavelength Divistion Multiplexers (DWDMs) to remove any parasitic light, effectively cleaning the pump and the seed laser signals, increasing the signal-to-noise ratio (SNR) at the idler frequencies. Afterwards, another DWDM is used to combine both signals and a PC is used to align the polarisation and maximise the coupling of light into the photonic chip. 
 The DWDMs consist of International Telegraphic Unit (ITU) channels in the telecom C-band. The pump and the seed laser are in channels 39 and 49 respectively while the stimulated idler generated in the MRR was filtered out using another DWDM at channel 29. The generated idler photons are recorded using a Finisar 1500S Optical Spectrum Analyser (OSA), allowing us to take JSI measurements in 20 - 40 second intervals with 1.2 pm (150 MHz) resolution, faster and more resolved than previous JSI measurement \cite{Grassani2016}. The whole process was automated using a computer. 

The device under test (DUT) is a photonic chip, based on the SOI platform and fabricated by a commercial foundry service (AMF Singapore). Specifically, the DUT consists of a micro racetrack resonator that is interferometrically coupled to a bus waveguide through a symmetric MZI. This gives rise to a tunable cavity Q-factor, which can be varied dynamically.

To obtain the purity from each measurement, we assume a flat  phase response and then, the JSA can be approximated by $\sqrt{JSI}$ \cite{Grassani2016}. 
Errors associated with the purity of the JSA were obtained by using the difference in estimated purity between the raw data, and data that was low-pass filtered using a 3x3 box filter. 
Nevertheless, the quoted purity is directly obtained from the unfiltered data. 

\begin{figure}[h]
\centering
\includegraphics[width=\linewidth]{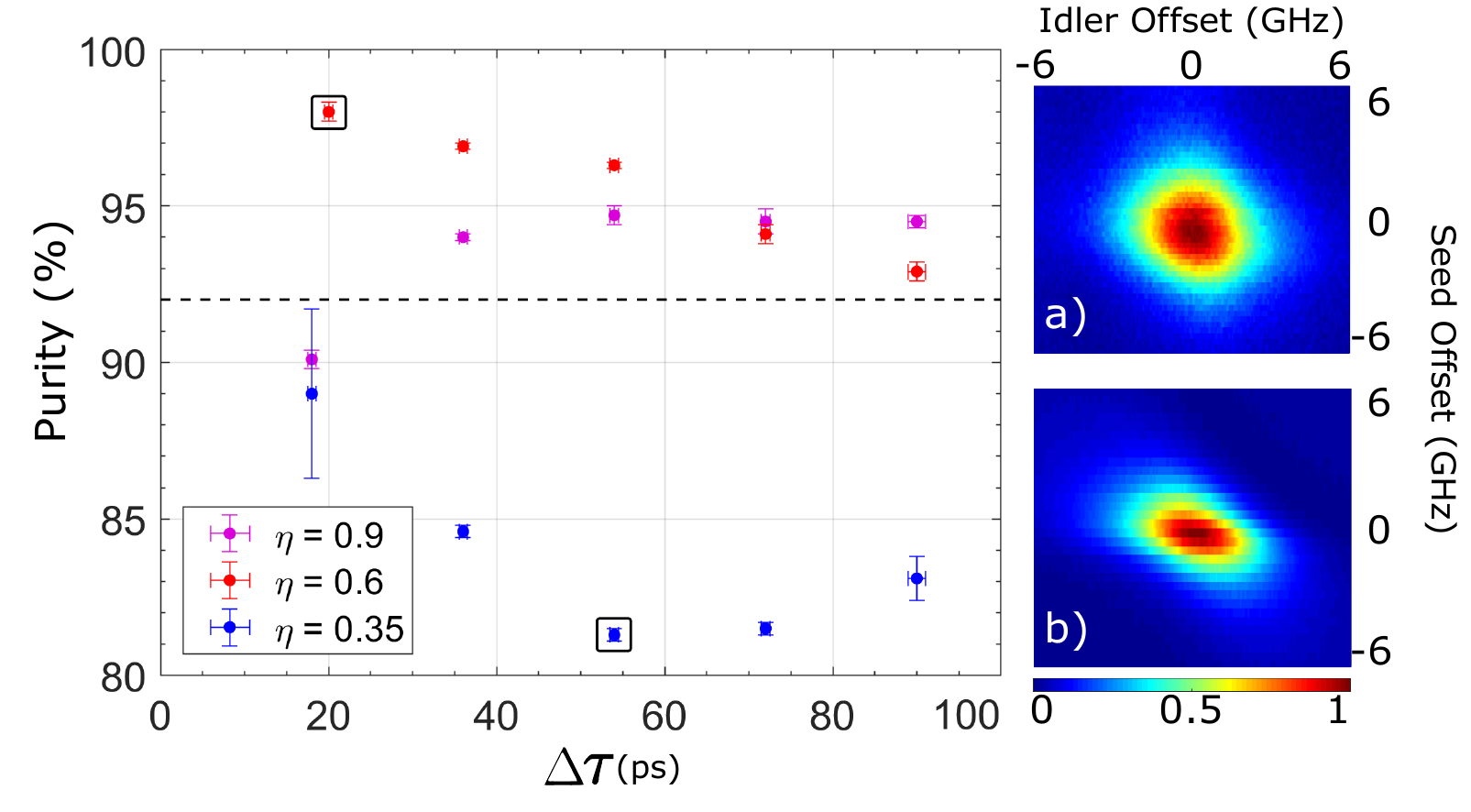}
\caption{The variation of purity with $\Delta\tau$ ($\tau_{p}$ = 420pm) for several values of $\eta$ ($\pm$ 0.02). JSIs are displayed for the two outlined data-points, with the dashed line representing 92\%. a) The JSI for the highest obtained purity (98.0 $\pm$ 0.3 \%). b) The JSI for the lowest obtained purity (81.3 $\pm$ 0.2 \%). }
\label{fig:Purity}
\end{figure}

Using the methods described above, we explored the parameter space available to us. First, we adjusted relative weighting of the pulses ($\eta$), as well as their temporal displacement ($\Delta\tau$); For $\eta$ = 0.6, and $\Delta\tau$ = 20ps we measured the purity of the JSI to be 98.0 $\pm$ 0.3 \% (Fig.~\ref{fig:Purity}a). To contrast this, for a $\Delta\tau$ of 54ps and an $\eta$ of 0.35; such that the 2nd pulse is strongest, we measured a purity reduction in the JSI down to 81.3 $\pm$ 0.2 (Fig.~\ref{fig:Purity}b). These results are consistent with the trends suggested in \cite{Christensen2018}, which are further mapped out in Fig.~\ref{fig:Purity}. The results suggest that by using this method not only can higher purities be reached, but the purity; and hence the degree of spectro-temporal entanglement can be fine-tuned independently from the resonator.

The purity of a resonator in the single pulse case is maximised when the linewidth of the pump resonance is small compared to the bandwidth of the pump laser, such that the intensity of the pump is flat across the resonance \cite{Helt2010}. As the difference between the resonance linewidth and the pump bandwidth decreases, the spectral purity of the emitted photon pairs starts to decrease.
It is possible to clarify whether we have saturated this ratio and the purity is maximal at 98\% or, alternatively, the purity is limited by the Q-factor of the cavity by exploring the dependence of purity on the Q factor of the cavity.

\begin{table}[htbp]
\centering
\caption{\bf Purity vs Quality-factor}
\begin{tabular}{ccccc}
\hline
Q ($\times10^3$) & $9.2 \pm 0.5$ & $12.3 \pm 0.5$ & $15.8 \pm 0.5$ & $19.6 \pm 0.5$ \\
\hline
Purity (\%) & $96.1 \pm 0.4$ & $97.2 \pm 0.5$ & $97.6 \pm 0.3$ & $97.9 \pm 0.6$ \\
\hline
\end{tabular}
\label{tab:Q-factor}
\end{table}

Table~\ref{tab:Q-factor} summarises measured purity values for different configurations of the MRRs MZI, effectively resulting in different Q-Factors (Q) for a pump bandwidth of 280pm.
The measured values for purity trend upwards as we increase the Q-factor of the cavity. However, the data also suggests that the measured purity (98\%) is limited by the relatively low Q ($\mathrm{2.5\times10^4}$) of the MRR cavity compared to suggested values in \cite{Christensen2018} or previous experiments \cite{Faruque2018}. It is nevertheless reasonable to expect higher Q-factors to provide purity values in excess of 99\%. 

\begin{figure}[h]
\centering
\includegraphics[width=\linewidth]{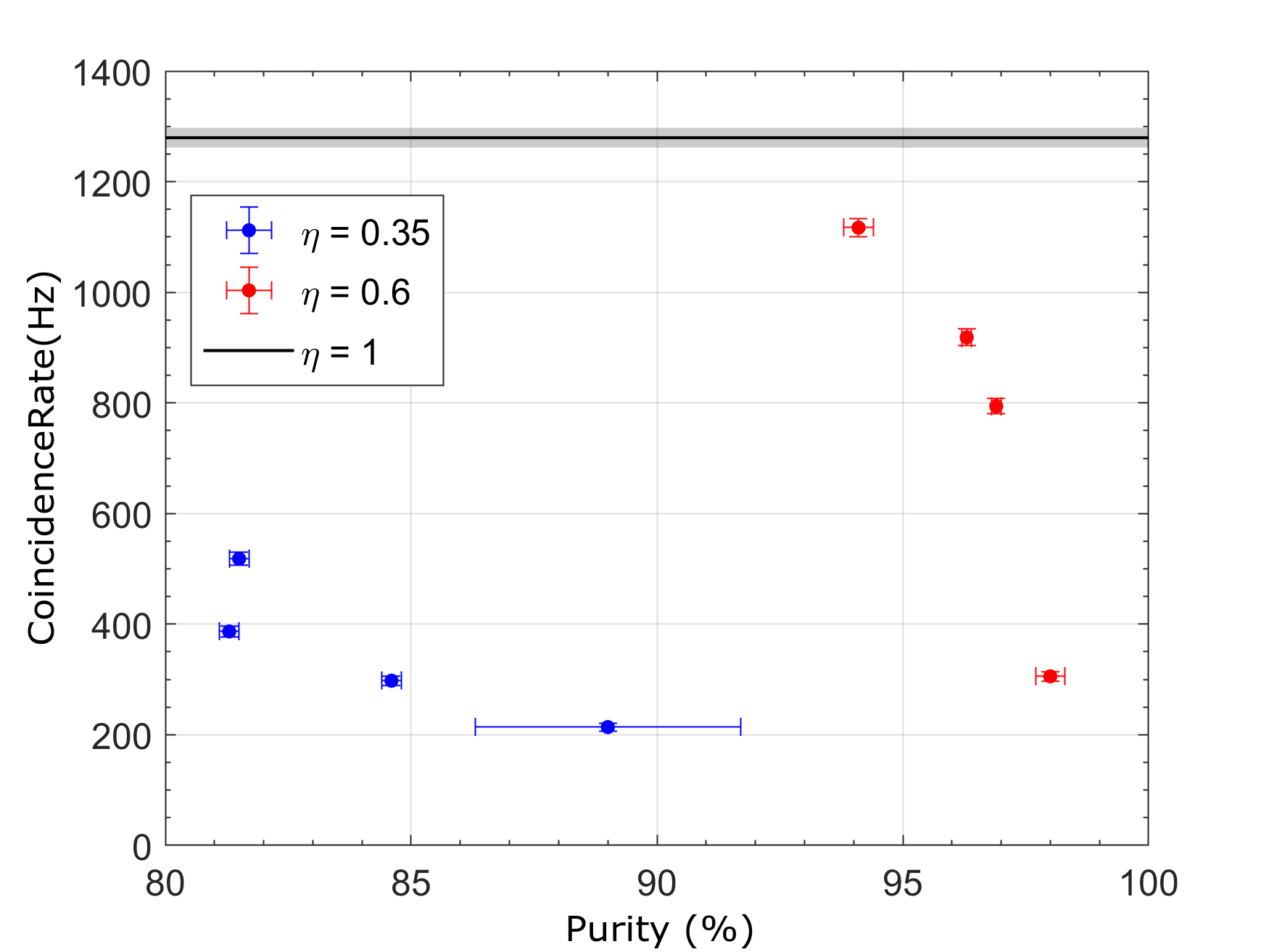}
\caption{The variation of coincidence rate with purity for an estimated pump power of 300$\mu W$ for varying $\eta$ ($\pm$ 0.02). 
}
\label{fig:Brightness}
\end{figure}

As discussed in \cite{Vernon:16} a common trade-off with MRRs tends to be between heralding efficiency and brightness. This describes the relationship between confinement for field enhancement, directly responsible for the brightness of the source, and photon pair extraction from the cavity, likewise related to the heralding efficiency; with both concepts depending on the Q-factor of the cavity.
It is possible to broaden the in-resonator pump spectrum compared to the signal and idler fields by changing the structure of the resonator as in \cite{Faruque2020}, gaining control over the resonance modulation at each frequency. Here we engineer the shape of the pulse inside the ring, leading to a change in the confinement of the pump and effectively achieving an equivalent outcome. Signal and idler photons are released independently of this without affecting the heralding rate. It is possible, therefore to measure the change in the coincidence rate as, when compared to the single pulse case, it approximates the relative brightness.
Coincidence rates for different values of $\eta$ are presented in Fig.~\ref{fig:Brightness} as a function of the purity. It can be seen that increasing the purity of the source will inevitably decrease its brightness for a specific $\eta$.

This additional trade-off with brightness seems to suggest that the purer the quantum state generated, the more finite those photon-pairs are. This begs the question as to whether it is more beneficial to spectrally filter a brighter signal, or to generate a weak but pure signal instead. The former leads to many unsuccessful operations, but pure throughput has the potential to be higher, whereas the latter points towards high fidelity operations at a low rate of occurrence.
A promising direction towards solving this problem involves the use of photonic molecules \cite{Borghi2019, Chuprina_2018} (multiple inter-coupled MRRs). It is not unreasonable to expect that by exploiting the inter-ring coupling within these molecules additional degrees of freedom could be manipulated to retain the brightness of less-pure sources.






In this paper we have experimentally demonstrated 98\% spectro-temporal purity of heralded single-photons generated from conventional MRRs by measuring the JSI using a time-delayed dual pulse configuration. We have varied the purity from 81.3 $\pm$ 0.2 \% to 98.0 $\pm$ 0.3 \% by adjusting the splitting ratio and the delay between the pulses, thus both below and above the fundamental limit of 92\% of single-pulse excitation of MRRs. Furthermore, we have observed how the increase in purity results in a trade-off with the photon-pair generation rate (i.e. brightness). By tuning the MZI-coupled MRR, we have also observed that the quality factor limits the maximum achievable purity. These observations qualitatively agree with the original dual-pulse proposal \cite{Christensen2018}. Our experimental demonstration enables silicon photonic devices with conventional MRR to produce photon pairs with fundamental spectro-temporal purity beyond the 92\% limit, by using off-the-shelf standard fibre-optic components. Furthermore, this approach potentially allows for multiple equivalent photon-pair sources to be driven in parallel with the same double-pulse source, paving the way for scalable heralded single photon sources. 

In the future, we will improve the stability of the dual-pulse configuration by integrating the interferometer on-chip. It is also imperative to investigate the trade-off between purity and brightness and its' limitations in a multi-photon experiment.

\medskip

\noindent\textbf{\Large Funding information.} This research was supported by the Engineering and Physical Sciences Research Council (EPSRC) grant no: EP-L024020/1 (Programme Grant); EP/M024458/1 (J.R. fellowship); EP/M013472/1 (Quantum Communications Hub); EP/M01326X/1 (Quantum Imaging Hub).
B.B. was supported by the Quantum Engineering Centre for Doctoral Training at Bristol, EPSRC Grant No. EP/L015730/1.
\medskip

\noindent\textbf{\Large Acknowledgements.} I.I.F. thanks Dr Damien Bonneau for the design of MZI coupled MRR. 
\medskip

\noindent\textbf{\Large Disclosures.} The authors declare no conflicts of interest.

\bibliographystyle{unsrt}
\bibliography{sample}



\end{document}